\documentclass[aps,pra,twocolumn]{revtex4-1}
\usepackage[T1]{fontenc}
\usepackage[latin1]{inputenc}
\usepackage{lmodern}
\expandafter\let\csname equation*\endcsname\relax
\expandafter\let\csname endequation*\endcsname\relax
\usepackage{amsmath}
\usepackage{amssymb}
\usepackage{graphicx}
\usepackage{xcolor}
\usepackage{natbib}
\usepackage{bm}
\usepackage{bbold}
\usepackage[mathscr]{euscript}
\usepackage[cal=boondoxo]{mathalfa}
\usepackage{comment}
\def\ii{{\rm i}}  
\def\GG{{\bf G}}

\def\db{\boldsymbol{\wp}}  
  
\def\Eb{{\bf E}}  
\def\rb{{\bf r}}

\def\bra#1{\mathinner{\langle{#1}|}}
\def\ket#1{\mathinner{|{#1}\rangle}}
\def\braket#1{\mathinner{\langle{#1}\rangle}}

\def\eq{\hat{\textbf{e}}_q} 
\def\eqp{\hat{\textbf{e}}_{q'}} 
\def\Gb{{\bf G}}  
  
\newcommand{\tj}[6]{ \begin{pmatrix}
   #1 & #2 & #3 \\
   #4 & #5 & #6 
  \end{pmatrix}}
\def\pM{\mathrel{\raise 2pt \hbox{\tiny(}\!\raise 1pt \hbox{+}\settowidth {\dimen03} {+}\hskip-\dimen03 \raise -2.4pt \hbox {$-$} \!\raise 2pt \hbox{\tiny)}}}

\begin{document}
\title{Optical waveguiding by atomic entanglement in multilevel atom arrays}
\author{A. Asenjo-Garcia$^1$}
\author{H. J. Kimble$^2$}
\author{D. E. Chang$^{3,4}$}
\affiliation{$^1$Physics Department, Columbia University, New York, NY 10027, USA}
\affiliation{$^2$Norman Bridge Laboratory of Physics MC12-33, California Institute of Technology, Pasadena, CA 91125, USA}
\affiliation{$^3$ICFO-Institut de Ciencies Fotoniques, The Barcelona Institute of Science and Technology, 08860 Castelldefels (Barcelona), Spain}
\affiliation{$^4$ICREA-Instituci\`o Catalana de Recerca i Estudis Avan\c{c}ats, 08015 Barcelona, Spain}

\date{\today}
\begin{abstract}
The optical properties of sub-wavelength arrays of atoms or other quantum emitters have attracted significant interest recently. For example, the strong constructive or destructive interference of emitted light enables arrays to function as nearly perfect mirrors, support topological edge states, and allow for exponentially better quantum memories. In these proposals, the assumed atomic structure was simple, consisting of a unique electronic ground state. Within linear optics, the system is then equivalent to a periodic array of classical dielectric particles, whose periodicity supports the emergence of guided modes. However, it has not been known whether such phenomena persist in the presence of hyperfine structure, as exhibited by most quantum emitters. Here, we show that waveguiding can arise from rich atomic entanglement as a quantum many-body effect, and elucidate the necessary conditions. Our work represents a significant step forward in understanding collective effects in arrays of atoms with realistic electronic structure.
\end{abstract}
\maketitle

Realizing efficient atom-light interactions is a major goal in quantum optics. Due to the intrinsically weak coupling between photons and atoms in free space, atomic ensembles have risen as one of the workhorses of the field, as the interaction probability with photons is enhanced due to the large number of atoms in the cloud~\cite{HSP10}. Atomic ensembles have broad potential applications, which include, among others, photon storage and retrieval~\cite{HHD99,FIM05,HSP10}, few-photon non-linear  optics~\cite{PWA13,PFL12,GTS14,TBS14}, and metrology~\cite{WJK10,LSV10,SKN12}. The fidelity of an atomic ensemble in carrying any of these applications is fundamentally limited by the so-called optical depth, which is a product of the interaction probability between a single atom and a photon in a given optical mode and the total number of atoms. While the important role of optical depth is ubiquitously stated in literature~\cite{GAF07,GAL07,CJG12,PKG16}, the underlying arguments in fact rely on one crucial assumption: that the atoms do not interact with each other and thus that photon emission happens at a rate given by that of single atoms. It is clear, however, that this approximation breaks down when atoms are close to each other, as photon emission is a wave phenomenon and interference and multiple scattering effects will be relevant at short distances.

In dense and ordered atomic arrays~\cite{LLK15,EBK16,BLL16,LBR16,BLL18,BSK17,BGP09,BPT10,WSF11}, strong constructive or destructive interference of light emitted by excited atoms allows one to exceed the fidelities predicted by these simple optical depth arguments in applications~\cite{AMA17}. For example, it has been theoretically shown that interference can impact communication and metrology applications: it enables both an exponential improvement in the fidelity of a quantum memory~\cite{AMA17,MMA18} and an improvement of the signal-to-noise ratio in optical lattice clocks~\cite{ORG13,HDC19}. More generally, interference in arrays can give rise to exotic phenomena~\cite{JR12,POR15,BGA16,FJR16,SR16,ZR10,ZR11,GGV18,GGV19}, which have no counterpart in disordered atomic gases. These include perfect reflection of light~\cite{BGA15,SWL17} or the existence of guided topological edge states of light in two-dimensional arrays~\cite{PBC17,BMA17}.

\begin{figure*}
\centerline{\includegraphics[width=0.85\linewidth]{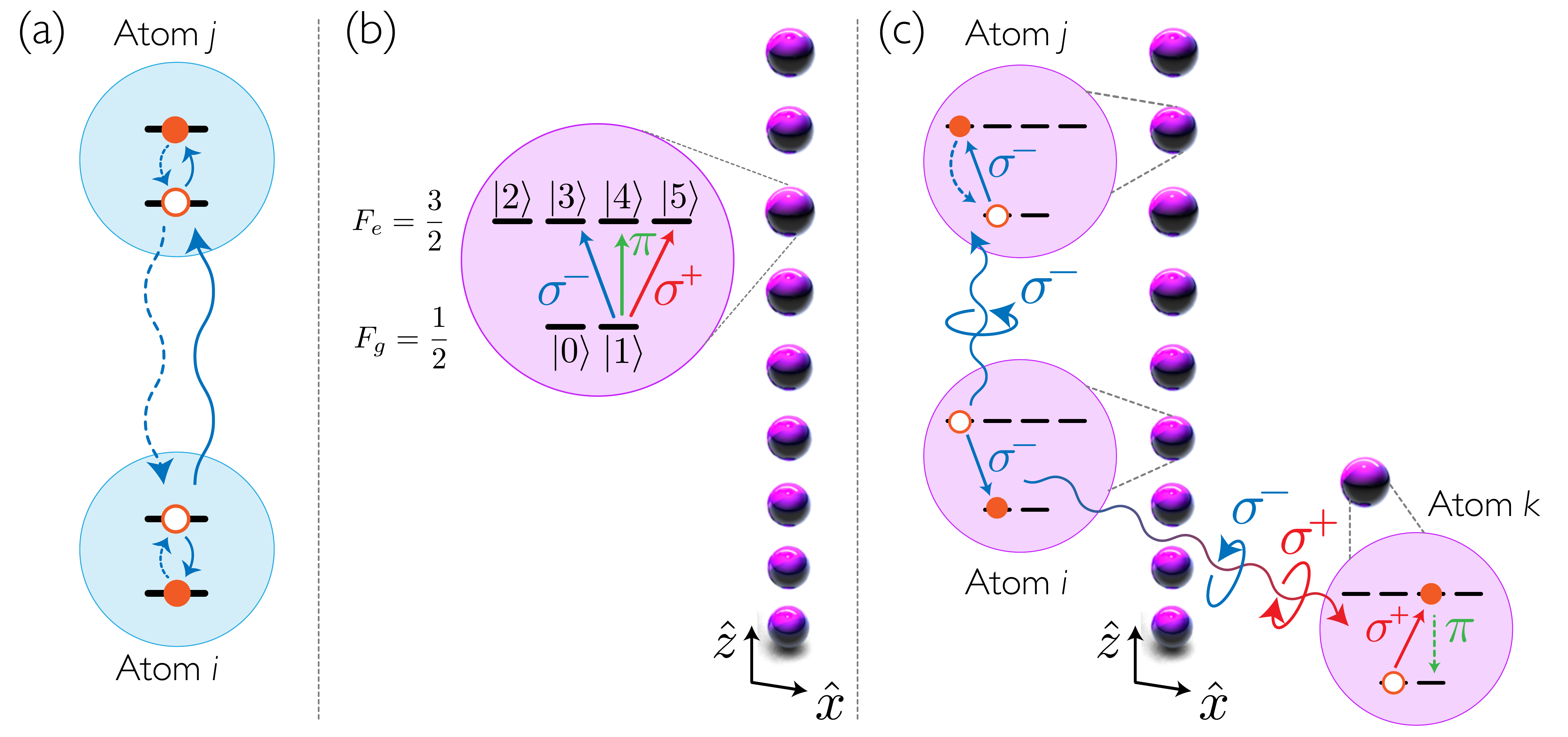}}
\caption{Illustration of the break-down of the two-level atom picture of dipole-dipole interactions, due to atomic hyperfine structure. \textbf{(a)} Illustration of photon-mediated interactions between two two-level atoms, with unique ground and excited states. \textbf{(b)} Schematic of a 1D array of multi-level atoms extended along the $z$-direction. The atoms considered have two ground states $\{\ket{0},\ket{1}\}$ with Zeeman quantum numbers $m_g=\{-1/2,1/2\}$, and four excited states $\{\ket{2},\ket{3},\ket{4},\ket{5}\}$, with quantum numbers $m_e=\{-3/2,-1/2,1/2,3/2\}$, respectively. The angular-momentum quantization axis lies parallel to the orientation of the chain. The transitions are coupled by photons of different polarization (depicted by different colors), such that $m_e-m_g=\{0,\pm1\}$, for polarizations $\{\pi,\sigma^\mp\}$, respectively.  \textbf{(c)} Illustration of the breakdown of a two-level subspace and subradiance. It is assumed that all atoms are initially in ground state $m_g=-1/2$, with the exception of atom $i$, which  decays from a stretched state emitting a photon. While atom $i$ necessarily ends up in state $m_g=-1/2$, the emitted photon does not have a spatially uniform polarization. In particular, in a geometry that is not purely 1D, the emitted photon could drive another atom $k$ out of the stretched two-level subspace (here illustrated by absorption of a $\sigma^{+}$ photon). Once atom $k$ is outside the two-level subspace, the excited state can decay into an unoccupied state (illustrated here by emission of a $\pi$-photon) at the rate of a single, isolated atom, which is not affected by collective effects.}\label{Fig1}
\end{figure*}

In these previous theoretical works, the atoms were assumed to have a unique electronic ground state. For two level atoms, and within the single-excitation manifold, multiple scattering enables a process where an excited atom $i$ can interact and exchange its excitation with another atom $j$ in its ground state [shown in Fig.~\ref{Fig1}(a)]. The resulting dynamical equations are exactly equivalent to $N$ classical polarizable dipoles interacting via their radiated fields. In particular, it is well known that ordered arrays of dielectric particles can support lossless guided modes~\cite{JMW95,BGS07,HBC15}. Within the context of infinite atomic arrays, waveguiding manifests itself in the form of perfectly ``subradiant'' single-excitation states with zero decay rate~\cite{SR16,AMA17}, a key idea underlying the previously proposed phenomena. In reality, though, most atoms display a rich hyperfine structure -- which arises from the coupling between the total electron and nuclear angular momenta -- and have more than one ground state. Given the growing body of theoretical and experimental literature about atomic arrays, it is critical to understand the underlying physics of collective optical phenomena for atoms with non-trivial internal structure. 

The complexity introduced by hyperfine structure is illustrated in Fig.~\ref{Fig1}, where one can see that light-mediated dipole-dipole interactions generally do not allow the atomic dynamics to be confined to a two-level subspace. In particular, even if atoms are initialized in such a subspace, emitted photons can drive other atoms out of the two level manifold, as photons do not have a uniform polarization in space. Once an atom is excited out of this subspace, the possibility to decay into unoccupied ground states cannot be suppressed by interference. Thus, even for a single excitation, the mechanism of subradiance, if it exists, could involve some many-body phenomenon. Indeed, the condition for subradiance to exist has already been investigated in the ``Dicke'' limit~\cite{HKO17}, where all atoms are located at a single point and thus effectively interact with a single, common electromagnetic mode. Interestingly, it was found that subradiance required a specific entanglement structure within the ground state manifold.

In this manuscript, we tackle the problem of collective effects in extended arrays of atoms with hyperfine structure. In particular, using a generalized ``spin model'' describing dipole-dipole interactions in the presence of hyperfine structure, we identify and analyze different classes of subradiant single-excitation states in a 1D atomic array. We find that the classical waveguiding effect still underlies the vast majority of subradiant states. Here, over large spatial regions, atoms in the array essentially live within a two-level subspace, interrupted by local ``defect'' states or domain walls that divide ``phase separated'' regions. However, we also describe a new, truly many-body mechanism, where waveguiding is enabled through rich, long-range entanglement within the ground-state manifold, and we elucidate the necessary conditions for its existence. These results are an important step forward in understanding collective effects in atoms with realistic electronic structure.

\section{Spin model and minimal toy atom}\label{sec2}
Here, we introduce a spin model to describe the photon-mediated quantum interactions between atoms. We will consider atoms whose electronic structure consists of a ground- and excited-state manifold, with total hyperfine angular momenta quantum numbers $F_g$ and $F_e$, respectively. A complete basis can be obtained by labeling states according to the projection of angular momentum along the $z$-axis, $\ket{F_{g/e}\,m_{g/e}}$, where $m_{g/e} \in[-F_{g/e},F_{g/e}]$ are Zeeman sublevels. We thus can describe the state of an atom by its quantum numbers, i.e., $\ket{F_e\,m_e}$ if it is excited or $\ket{F_g\,m_g}$ if it is in the ground state manifold. The ground and excited states couple to light via well-defined selection rules, such that $m_e=m_g+q$, with $q=\{0,\pm1\}$ denoting the units of angular momentum that can be transferred by a photon. We can define an atomic raising operator that depends on $q$ as
\begin{equation}
\hat{\Sigma}^\dagger_{iq}=\sum_{m_g=-F_g}^{F_g} C_{m_g ,q}\hat{\sigma}^i_{F_e m_g-q,F_g m_g},
\end{equation}
where $\hat{\sigma}^i_{F_e m_g-q,F_g m_g}=\ket{F_e \,m_g-q}_i\bra{F_g\, m_g}_i$ is the atomic coherence operator between the ground and excited states of atom $i$. This operator conveniently groups all the possible transitions which transfer $q$ units of angular momentum along $z$. These are weighted by the Clebsch-Gordan coefficients
\begin{align}
C_{m_g,q}=(-1)^{F_g-m_g}\tj {F_g} 1 {F_e} {-m_g} q {m_g-q},
\end{align} 
written here in terms of a Wigner 3j-symbol, which reflect the different strengths that the possible transitions couple to light of a given polarization.

\textit{Spin model for multilevel atoms}-- Intuitively, the interaction with light allows for processes of photon-mediated emission and re-absorption, where the excitation of one atom decays and another is excited. We describe such dynamics by means of a spin model~\cite{GW96,BI12,AHC17,GH1982,EKM06,KEK07,MAL18,HOR19} where we integrate out the photons and find an all-atomic density matrix that only depends on the internal degrees of freedom of the atoms. Specifically, the evolution of the atomic density matrix $\hat{\rho}_{\rm A}$ obeys $\dot{\hat{\rho}}_{\rm A}=-(\ii/\hbar)\,[\mathcal{H},\hat{\rho}_{\rm A}]+\mathcal{L}[\hat{\rho}_{\rm A}]$, where $\mathcal{H}$ is the Hamiltonian, and $\mathcal{L}[\hat{\rho}_{\rm A}]$ is the Lindblad operator.  For atoms with hyperfine structure and resonance frequency $\omega_0$, these operators read 
\begin{subequations}\label{fullham}
\begin{equation}
\mathcal{H}=\hbar\sum_{i,j=1}^N\sum_{q,q'=-1}^1J_{ijqq'}\hat{\Sigma}^{\dagger}_{iq}\hat{\Sigma}_{jq'},
\end{equation}
\begin{align}\label{lindblad}
\mathcal{L}[\rho]=\sum_{i,j=1}^N\sum_{q,q'=-1}^1\frac{\Gamma_{ijqq'}}{2}\left(2\hat{\Sigma}_{jq'}\rho\hat{\Sigma}^{\dagger}_{iq} -\hat{\Sigma}^{\dagger}_{iq}\hat{\Sigma}_{jq'}\rho\right.\\\nonumber
 \left.-\rho\hat{\Sigma}^{\dagger}_{iq}\hat{\Sigma}_{jq'}\right),
\end{align}
\end{subequations}
where we have defined the polarization-dependent spin-exchange and decay rates as
\begin{subequations}
\begin{equation}
J_{ijqq'}=-\frac{\mu_0\omega_0^2}{\hbar}|\boldsymbol{\wp}|^2 \,\,\eq\cdot\text{Re}\,\Gb (\rb_i,\rb_j,\omega_0)\cdot\eqp^*,
\end{equation}
\begin{equation}
\Gamma_{ijqq'}=\frac{2\mu_0\omega_0^2}{\hbar}|\boldsymbol{\wp}|^2 \,\,\eq\cdot\text{Im}\,\Gb (\rb_i,\rb_j,\omega_0)\cdot\eqp^*,
\end{equation}
\end{subequations}
with $\db=\braket{F_g||e\hat{\rb}||F_e}$ being the reduced matrix element associated with the transition. In the above equations, $\hat{\textbf{e}}_{\pm 1}=\mp\frac{1}{\sqrt{2}}(\hat{x}\pm\ii\hat{y})$, and $\hat{\textbf{e}}_0=\hat{z}$ are spherical basis vectors. Physically, the spin-exchange and decay rates are proportional to the classical field amplitude projected along polarization $q$ at position $\rb_i$, due to a classical oscillating dipole of polarization $q'$ at $\rb_j$. Naturally, they are given in terms of the free-space electromagnetic Green's tensor, $\GG(\rb_i,\rb_j,\omega_0)\equiv\GG(\rb_{ij},\omega_0)$, with $\rb_{ij}=\rb_i-\rb_j$, which reads
\begin{align}
\GG(\rb,\omega_0) = \frac{e^{\ii k_0 r}} 
{4\pi k_0^2 r^3} \left[(k_0^2 r^2+\ii k_0 r -1) \mathbb{1} \right.+ \notag \\
\qquad{}\left. + (-k_0^2 r^2 -3\ii k_0 r + 3) \frac{\rb \otimes \rb}{r^2}  \right], \label{Greens_def}
\end{align}
where $r\equiv|\rb|$ and $k_0=2\pi/\lambda_0=\omega_0/c$ is the wave number corresponding to the atomic transition energy. Note that the Green's function $\GG_{\alpha\beta}$ is a tensor quantity ($\{\alpha,\beta\}=\{x,y,z\}$), as both the electromagnetic field and the atomic transition have specific polarizations.  These equations are derived within the Markovian approximation, which is highly justified for our system under consideration~\cite{CJG12,SCC15,GRL16}. 

The dynamics under the master equation can analogously be described in the quantum jump formalism~\cite{GZ04}, where the last two terms in the parentheses of Eq.~\eqref{lindblad} are combined with $\mathcal{H}$ to form a non-Hermitian Hamiltonian that characterizes the deterministic evolution. Within this formalism, the effective non-Hermitian Hamiltonian is readily given by
\begin{equation}\label{heffqq}
\mathcal{H}_\text{eff}=\hbar\sum_{i,j=1}^N\sum_{q,q'=-1}^1\left(J_{ijqq'}-\ii\frac{\Gamma_{ijqq'}}{2}\right)\hat{\Sigma}^{\dagger}_{iq}\hat{\Sigma}_{jq'}.
\end{equation}
While this equation is quite general, we now highlight several points, which will be key to understanding collective decay: 
\begin{itemize}
\item First, for a single, isolated atom, we obtain the total decay rate from excited sublevel $m_e$ as $\Gamma_0=\sum_{q}\Gamma_{iiqq}C_{m_e+q,q}^2$, which is equal for all $m_e$. The branching fraction into a specific ground state $m_g=m_e+q$ is simply given by $C_{m_e+q,q}^2$. 

\item Second, the form of $\mathcal{H}_\text{eff}$ illustrates that, in general, lowering one atom angular momentum projection by $q$ does not necessarily imply that another atom's projection is raised by $q$ ($q' \neq q$), as $q$ refers to the dipole matrix element and not the global polarization of the emitted photon [see Fig.~\ref{Fig1}(c)]. Thus, a system initialized in a two-level subspace does not generally remain in that space under coherent dynamics. In particular, even if the initially involved excited state decays to a unique ground state, a different excited state can become populated later in time [see atoms $i$ and $k$ in Fig.~\ref{Fig1}(c)]. We note that one way to suppress such undesired excitation pathways (e.g., the $\sigma^{+}$ transition of atom $k$) is by applying large Zeeman shifts. While this might indeed constitute a practical way to maintain the desired effects of subradiance to some extent given hyperfine structure, our focus here will be on identifying new mechanisms for subradiance, which fundamentally persist in the presence of multiple pathways.

\item Finally, we consider the situation where an excited atom has an allowed decay channel into a ground state that is unoccupied by any atoms [illustrated by atom $k$ in Fig.~\ref{Fig1}(c) following absorption of a $\sigma^+$ photon]. In that case, there is no interference; the excited state decays into that ground-state level with the full strength of a single, isolated atom. Presumably, any mechanism for strong subradiance must then prevent this process from occurring. 
\end{itemize}

It will also be useful to calculate the emitted quantum field associated with any given atomic state. It can be shown that the positive frequency component reads
\begin{equation}\label{efield}
\hat{\Eb}^+(\rb)=\mu_0\omega_0^2\sum_{j=1}^N\sum_{q=-1}^1\Gb\,(\rb,\rb_j,\omega_0)\cdot\hat{\textbf{e}}^*_{q}\,\,\wp\,\hat{\Sigma}_{jq}.
\end{equation}
From the previous equation, we can find the negative-frequency component $\hat{\Eb}^-(\rb)$ by taking the Hermitian conjugate of $\hat{\Eb}^+(\rb)$. The field also in principle contains a vacuum noise component~\cite{AMA17}, which will not affect our quantities of interest and is thus not explicitly written.

\begin{figure}
\centerline{\includegraphics[width=\linewidth]{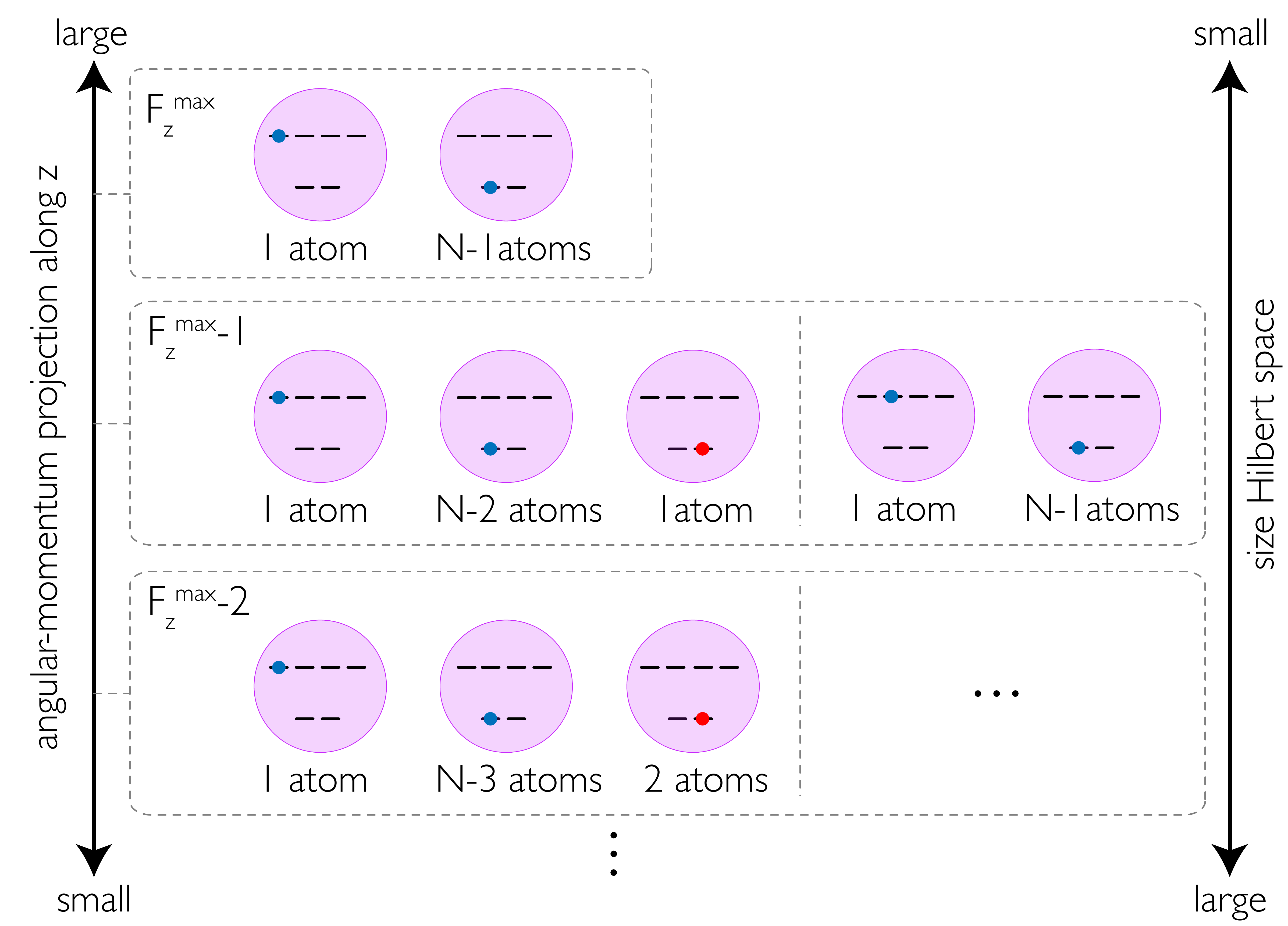}}
\caption{Classification of (single-excitation) basis states within each subspace of conserved angular momentum projection $F_z$ along the z-axis. For $|F_z|=F_z^\text{max}$, the states live within a two-level subspace, where one $(N-1)$ atoms occupy the excited (ground) states of minimum $m_e (m_g)$. As $|F_z|$ is lowered, one type of basis state consists of replacing some number of the ground state atoms of minimum $m_g=-1/2$ with atoms (shown in red) with $m_g=+1/2$. These basis states containing ``defect'' atoms pre-dominantly compose subradiant states, as described in the main text.}\label{Fig2}
\end{figure}
\textit{One-dimensional chain}-- From now on, we will focus on the problem of a 1D chain. We will see that this system both encodes the classical waveguiding effect previously identified for two-level atoms, and allows for new highly correlated subradiant states due to multi-level structure. Moreover, 1D allows for numerics on relatively large systems (not dominated by boundaries), and we can take advantage of additional conserved quantities that allow us to simplify the problem. For the sake of simplicity, we choose the direction of the chain to align to the quantization axis of the atoms (i.e., along $z$). This guarantees the conservation of polarization of the emitted photons, as $\GG_{\alpha\beta}=0$ along the axis if $\alpha\neq\beta$ (i.e., $\GG_{\alpha\beta}$ is a diagonal tensor in the polarization indices). Therefore, the only non-zero interactions are those preserving polarization (i.e., $q= q'$) and the effective Hamiltonian simplifies to
\begin{equation}
\mathcal{H}_\text{eff}=\hbar\sum_{i,j=1}^N\sum_{q=-1}^1\left(J_{ijq}-\ii\frac{\Gamma_{ijq}}{2}\right)\hat{\Sigma}^{\dagger}_{iq}\hat{\Sigma}_{jq}.
\end{equation}
where $J_{ijq}\equiv J_{ijqq}$ and $\Gamma_{ijq}\equiv\Gamma_{iiqq}$. 

Until now, our spin model captures the dynamics of atoms with any kind of hyperfine structure. Hereafter, we focus on what we consider to be the minimal toy model that captures all the relevant physics: 6-level atoms with $F_g=1/2$ and $F_e=3/2$ [see Fig.~\ref{Fig1}(b)]. This specific hyperfine structure displays closed transitions, where an excited state with maximum angular momentum projection $m_e=\pm 3/2$ can only decay to only one ground state, also with maximum angular momentum projection $m_g=\pm 1/2$ in the ground state manifold by emitting a circularly-polarized photon ($q=\mp1$). It also exhibits excited states with $m_e=\pm 1/2$ where decay into two different ground states $m_g=\pm 1/2$ is allowed, which involve emission of photons of different polarizations ($q=\pm 1$ or $q=0$). To simplify notation, in what follows we will label ground states with $m_g=\{-1/2, 1/2\}$ as $\{\ket{0},\ket{1}\}$, respectively, and excited states with $m_e=\{-3/2, -1/2, 1/2, 3/2\}$ as $\{\ket{2},\ket{3},\ket{4},\ket{5}\}$, respectively [see Fig.~\ref{Fig1}(b)].

The conservation of angular momentum projection along the direction of the chain allows us to diagonalize the Hamiltonian by blocks with well-defined $F_z=\sum_{i=1}^N m_i$, where the sum over Zeeman sublevels includes both ground and excited states (see Fig.~\ref{Fig2}). Note that subspaces with equal magnitude of angular-momentum projection but opposite sign (i.e., $\pm F_z$) display the same energy and decay spectra  (i.e. the physics is identical if we simultaneously flip $m_g$ to $-m_g$ and $m_e$ to $-m_e$). In what follows, we study single-excitation subradiant states of various angular momentum manifolds, starting by maximum $|F_z|$ (as we will show, the most ``classical'' manifold) and ending by $F_z=0$ (the most complex, and where subradiant states enabled by rich entanglement live). 

Motivated by previous work on 1D and 2D arrays with simple atomic structure~\cite{AMA17}, here we seek to elucidate the properties of eigenstates $\ket{\psi_\xi}$ of $\mathcal{H}_\text{eff}$ in the single-excitation manifold. In particular, such states will have complex eigenvalues $\omega_\xi=J_\xi-i\Gamma_{\xi}/2$ characterizing the energy shifts and decay rates. Of specific interest will be the identification of states where the decay rate approaches zero as $N\rightarrow\infty$, which implies that the states decouple from radiation fields and correspond to guided modes; and to elucidate the properties of the eigenstates that enable the waveguiding phenomenon. Importantly, while for simple two-level atoms, the Hilbert space of the single-excitation manifold increases as $\sim N$ and encodes classical linear optics, here, the manifold size increases exponentially, raising the possibility for waveguiding through entanglement.

\section{Two-level subradiance: $|F_z|=F_z^\text{max}$}\label{sec3}
\begin{figure}[h!]
\centerline{\includegraphics[width=\linewidth]{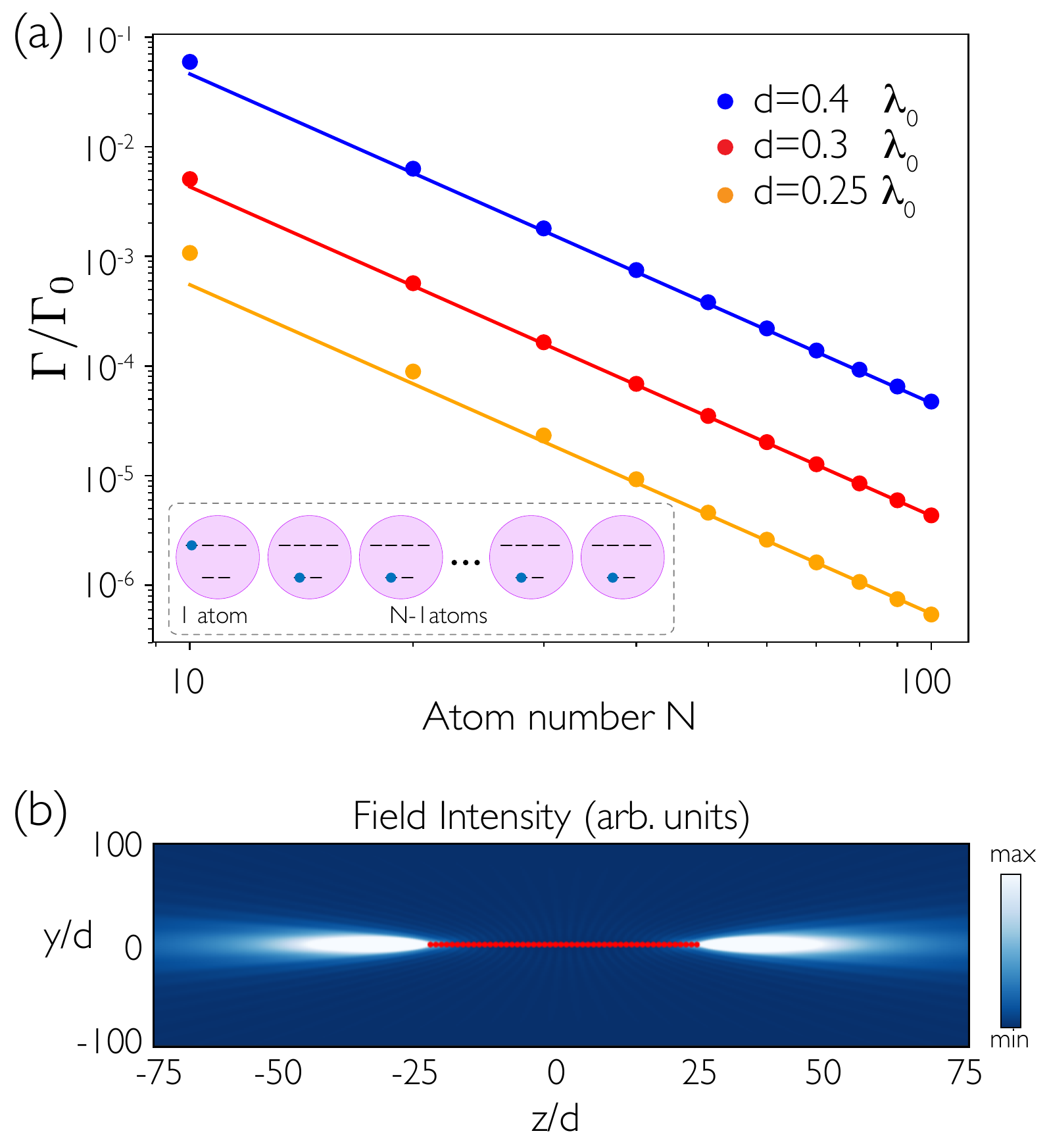}}
\caption{Properties of the most subradiant eigenstate in the $|F_z|^\text{max}$ manifold. \textbf{(a)} Decay rate $\Gamma$ of the most subradiant eigenstate vs. atom number, for different lattice constants. The continuous lines are guides to the eye and scale as $\Gamma/\Gamma_0\sim1/N^3$. The inset illustrates the kind of states that appear in this manifold (where one atom is in $\ket{2}$ and all the others in $\ket{0}$). \textbf{(b)} Field intensity (arbitrary units) emitted by the most subradiant mode in a chain of $N=50$ atoms. The field is largely evanescent transverse to the bulk of the chain, while most of the energy is radiated out through scattering at the ends of the chain, as expected for an optical waveguide. Red circles denote  atomic positions.} \label{Fig3}
\end{figure}

In each subspace $F_z$, it is first helpful to consider the full set of possible basis states. For example, within the single-excitation manifold, the maximum allowed value of $F_z$ is given by $F_z^\text{max}=\pm |F_e+(N-1)F_g|=\pm (N/2+1)$, which is achieved when one atom is excited in the state $\ket{2}$, with $m_e=-3/2$, while all other atoms are in the ground state $\ket{0}$, with $m_g=-1/2$. These states are connected through a $\sigma^-$ transition (this is, $q=1$). Since $\mathcal{H}_\text{eff}$ is block-diagonalizable, $|F_z|=F_{z}^\text{max}$ thus corresponds to an effective two-level atom subspace.

Numerically, we diagonalize $\mathcal{H}_\text{eff}$ within this subspace as a function of atom number $N$ and for selected different lattice constants $d$. For each $N$ and $d$, we then find the eigenstate with the minimum decay rate $\Gamma$, which we plot in Fig.~\ref{Fig3}(a). In agreement with our previous results~\cite{AMA17}, we find subradiant states if the interatomic distance $d$ is such that $d<\lambda_0/2$, $\lambda_0$ being the wavelength of the resonant transition. The decay rate of the most subradiant eigenstate scales as $\Gamma/\Gamma_0\sim1/N^3$, as also found previously for two level atoms. We calculate the field intensity radiated by the most subradiant eigenstate $\ket{\psi}$ by numerically finding the expectation value of  the intensity operator $I(\rb)=\braket{\psi|\hat{\Eb}^-(\rb)\cdot\hat{\Eb}^+(\rb)|\psi}$, according to Eq.~\eqref{efield}. The intensity pattern  in Fig.~\ref{Fig3}(b) demonstrates strong emitted intensity as the ends of the chain and absence of intensity in the middle, as would be expected for a finite-size waveguide. 

As previously remarked, the guiding effect is entirely classical. Because the ground state is unique and there is only one excitation in the system (i.e., $N$ degrees of freedom), one can alternatively deduce this state by considering $N$ coupled harmonic oscillators. In that case, the stationary states under full master equation evolution are coherent states $\ket{\psi_{\xi}^{ho}}=\ket{\alpha_1 e^{-i\omega_\xi t}}\ket{\alpha_2 e^{-i\omega_\xi t}}...\ket{\alpha_N e^{-i\omega_\xi t}}$. Here, the coherent state amplitude $\alpha_i\propto c_i$ of harmonic oscillator $i$ is proportional to the spin wavefunction amplitude of the corresponding single-excitation spin eigenstate $\ket{\psi_{\xi}}=\sum_i c_i \sigma_{20}^{i} \ket{0}^{\otimes N}$ found earlier. These coherent states amplitudes evolve in time as $e^{-i\omega_\xi t}$ with a frequency $\omega_\xi$ corresponding to the spin state eigenvalue. These harmonic oscillator states physically describe classical resonant dipole arrays, interacting with each others' radiated fields. In particular, it is well-known that an infinite array can support guided modes. Starting from such states, we can apply an operator $\mathcal{P}$ that projects these states into the single-excitation manifold, which then exactly reproduces the single-excitation spin eigenstates $\ket{\psi_\xi}$. This implies that the resulting physics can be understood purely classically -- even if the spin eigenstates themselves are technically entangled -- as the entanglement simply arises from the projection at the end. As we will see later on, there are entangled states that do not support any analogy with those of classical dipoles.

\begin{figure*}
\centerline{\includegraphics[width=\linewidth]{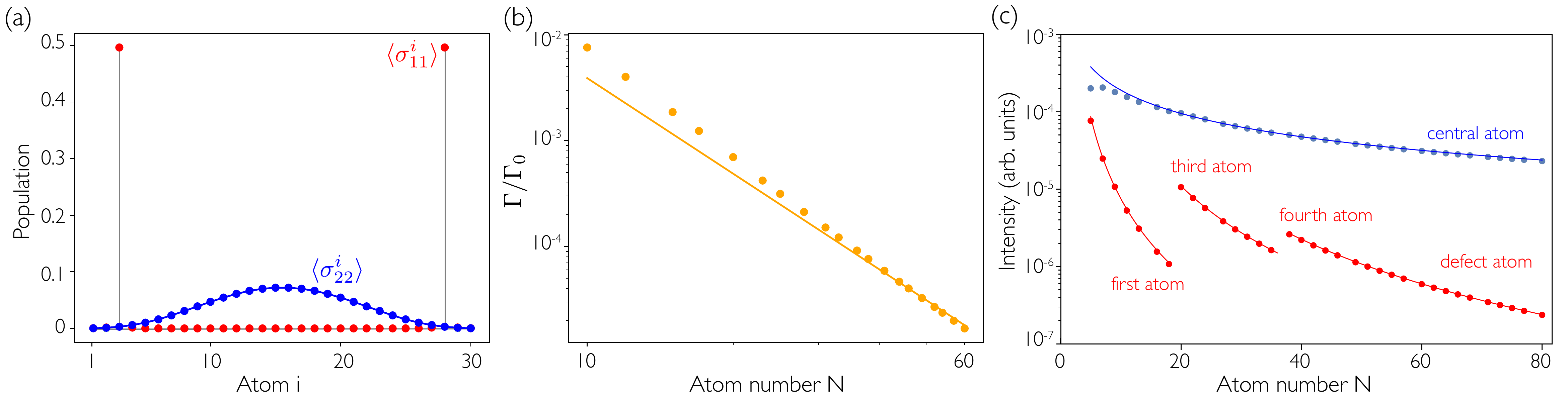}}
\caption{Defect states in $|F_z|=|F_z^\text{max}|-1$. {\bf (a)} Spatial profile of the populations of levels $\ket{1}$ (red) and $\ket{2}$ (blue) in the most subradiant eigenstate of a chain of $N=30$ atoms. The population of $\ket{3}$ is negligible in all atoms. The continuous lines are guides to the eye. {\bf (b)} Decay rate of the most subradiant eigenstate vs. atom number $N$. The line is a guide to the eye that scales as $\Gamma/\Gamma_0\sim 1/N^3$. {\bf (c)} Field intensity (in arbitrary units) at the central and defect atoms. The lines are guides to the eye, and follow $I\sim 1/N$ (blue) and $I\sim1/(N-n)^3$ (red) with $n=\{1,3,4\}$, where $n$ represents the position of the atomic defect. For all plots, $d=0.3\lambda_0$.} \label{Fig4}
\end{figure*}

The waveguiding concept can be additionally confirmed by considering an infinite chain. In this case, we can diagonalize the Hamiltonian using spin waves with well-defined momentum $k$ along $z$, and readily find $\mathcal{H}_\text{eff}=\sum_{q=-1}^1 \tilde{\mathcal{H}}_\text{q}$, where 
\begin{equation}\label{hq}
\tilde{\mathcal{H}}_\text{q}=\hbar\sum_{k}\left(J_{k,q}-\ii\frac{\Gamma_{k,q}}{2}\right)\hat{S}^{\dagger}_{k,q}\hat{S}_{k,q},
\end{equation}
with $\hat{S}^\dagger_{k,q}= N^{-1/2} \sum_{j}  e^{\ii k d j} \hat{\Sigma}^\dagger_{jq}$ and
\begin{subequations}\label{jjgg}
\begin{equation}
J_{k,q}=-\frac{\mu_0\omega_0^2}{\hbar}|\boldsymbol{\wp}|^2 \,\,\eq\cdot\text{Re}\,\tilde{\Gb}  (k,\omega_0)\cdot\eq^*,
\end{equation}
\begin{equation}
\Gamma_{k,q}=\frac{2\mu_0\omega_0^2}{\hbar}|\boldsymbol{\wp}|^2 \,\,\eq\cdot\text{Im}\,\tilde{\Gb} (k,\omega_0)\cdot\eq^*.
\end{equation}  
\end{subequations}
In the above equations, $\tilde{\GG} (k) = \sum_{j} e^{-\ii k d j} \GG (z_j)$ is the discrete Fourier transform of the free-space Green's tensor. Eigenstates in the single-excitation manifold can be generated by applying a spin raising operator $S^{\dagger}_{k,1}$ to the product ground state $\ket{0}^{\otimes N}$. For $k>\omega_0/c$ -- such that the spin wavevector exceeds the wavevector of free-space radiation -- one finds that $\Gamma_{k,q}=0$, indicating the decoupling of the spin wave from radiation fields and thus the guided nature of these excitations. We note that the ability to restrict dynamics to a two-level subspace is unique to a 1D chain, as in higher dimensions the Hamiltonian of Eq.~\eqref{heffqq} is not restricted to $q=q'$. This motivates the deeper investigation of subradiance in subspaces of other total $F_z$, to find a true many-body mechanism.

\section{Defect states: $|F_z|=|F_z^\text{max}|-1$}

Reducing the angular momentum projection by one unit increases the complexity of the problem. This is manifest in the larger size of the Hilbert subspace, of dimension $dim(\mathcal{H})=N^2$. The basis states are of two types: (i) ``defect'' states where one atom is excited in $\ket{2}$ ($m_e=-3/2$), $N-2$ atoms are in $\ket{0}$ ($m_g=-1/2$) (reminiscent of the two-level subspace of the previous sub-section), and one ``defect'' atom is in the ground state $\ket{1}$ ($m_g=1/2$); (ii) states where all atoms except one are in the ground state $\ket{0}$, and the excited state involves level $\ket{3}$ ($m_e=-1/2$), a state without maximal angular momentum projection. From the considerations of Sec. II, it is already clear that any highly subradiant states must exhibit the following features. First, because the basis states containing excited state $\ket{3}$ have an allowed transition (via a $q=1$ photon) to state $\ket{1}$, which is unoccupied, such states will decay into $\ket{1}$ at the full strength $\Gamma_0/3$ of a single, isolated atom. Thus, these basis states must constitute a vanishingly small weight of a subradiant state. Second, for basis states composed of a single atom $j$ in state $\ket{1}$, one can conclude that coherent interactions with an atom $i$ in state $\ket{2}$ allow for an exchange $\ket{2_i\,1_j}\rightarrow\ket{0_i\,3_j}$, while we argued already that state $\ket{3}$ is undesirable. Thus, a subradiant state must find some mechanism to ``hide'' the atom in state $\ket{1}$ from the dynamics. 

This intuition agrees with our numerical findings. For the most subradiant eigenstates, the population of state $\ket{3}$ decreases as $\sum_j\braket{\hat{\sigma}_{33}^j}\sim 1/N^3$, i.e., in the thermodynamic limit the most subradiant eigenstate is mostly formed by ``defect'' states and is of the form
\begin{equation}
\ket{\psi}=\sum_{i,j=1}^N  c_{ij}\,\hat{\sigma}_{20}^i\hat{\sigma}_{10}^j\ket{0}^{\otimes N}.
\end{equation}

Figure~\ref{Fig4}(a) shows the spatial profile of the population in levels $\ket{1}$ and $\ket{2}$ of the most subradiant eigenstate of a chain of $N=30$ atoms. The population in the excited state $\ket{2}$ is spatially smooth, as it is distributed among all atoms. On the other hand, the population in the defect level $\ket{1}$ is localized in a superposition between both edges of the chain (in particular, the defect atom is the third one, counting from both edges). The population of level $\ket{3}$ is spatially correlated with that of $\ket{1}$, as $\braket{\sigma_{33}^j}\propto \braket{E^-(\rb_j)E^+(\rb_j)} \braket{\sigma_{11}^j}$  (see SI). The population in this level is also localized around the third atom, but negligible in the scale of the figure. As shown in Fig.~\ref{Fig4}(b), the decay rate of the most subradiant state with atom number scales as $\Gamma/\Gamma_0\sim 1/N^3$. Figure~\ref{Fig4}(c) shows the scaling of the field intensity at the position of the central atom of the chain versus that at the defect atom. The latter scales as $I\sim1/N^3$. This confirms our previous intuition, that the defect atom is efficiently ``hidden'' from the rest of the chain. In particular, the low intensity seen by the defect atom ensures that it cannot be efficiently excited to level $\ket{3}$, and that the large decay rate of this excited state does not destroy subradiance. The precise spatial location of the defect atom depends on microscopic details, and changes depending on the length of the chain (it appears just at the edge for short chains and in atoms further away from the boundaries for longer chains), but the precise location does not qualitatively alter the physics (see SI). 

Furthermore, when $F_z=|F_z^{max}|-n_d$ is further reduced, (with $n_d\ll |F_z^{max}|$), the most subradiant states are of similar character, with long chains of ``two-level'' atoms interrupted by $n_d$ defects, whose potentially detrimental effects are reduced as they are largely decoupled (see SI). Importantly, the character of such states, dominated by the waveguiding of effective two-level atoms, do not yet point to any non-trivial mechanism for subradiance.

\begin{figure*}
\centerline{\includegraphics[width=\linewidth]{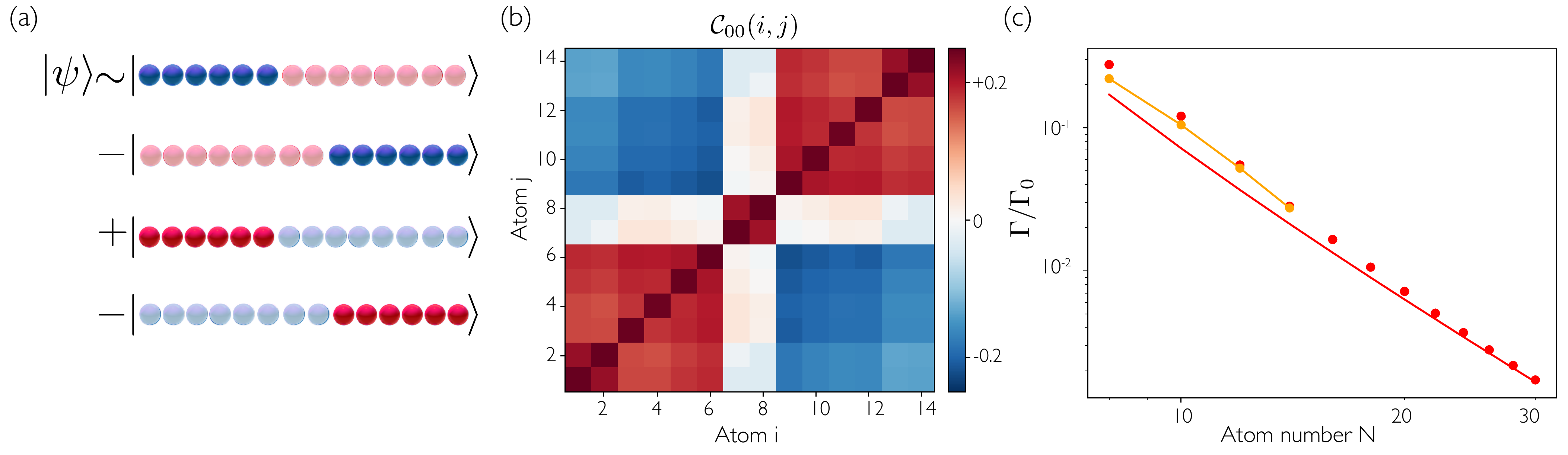}}
\caption{Phase separated subradiant states in $F_z=0$. {\bf (a)} Sketch of the most subradiant eigenstate. Blue (red) circles represent the phase with one excited atom in $\ket{2}$ ($\ket{5}$) and $N/2-2$ atoms in $\ket{0}$ ($\ket{1}$). The faded circles represent the ``inactive'' or defect phase. {\bf (b)} Evidence of phase separation in a chain of $N=14$ atoms through correlations between $\ket{0}$ states at atoms $i,j$. {\bf (c)} Scaling of decay rate with atom number for phase separated states (orange line). In red, scaling of most subradiant state of a chain of $N/2-1$ atoms in $|F^\text{max}_z|$. The continuous line is a guide to the eye showing a scaling of $\Gamma/\Gamma_0\sim1/(N/2-1)^3$. For the middle and right plots, $d=0.3\lambda_0$.} \label{Fig5}
\end{figure*}

\section{Phase-separated and symmetric states: $F_z=0$}
For an even number of atoms, there is a subspace with zero angular momentum projection. This manifold is the largest one, with dimension that scales exponentially with atom number. In this manifold, we have approximately the same number of atoms in each of the degenerate ground states, allowing for many different combinations. We are able to find two different kinds of subradiant states: phase separated states and symmetric states. In the following, we study these two families in detail. 

\subsection{Phase-separated states}
Decreasing angular momentum increases the number of defects, which tend to localize at the chain edges. In $F_z=0$, approximately half of the chain is made of defects, and phase separation and domain walls emerge, which we discuss in this section. Phase-separated states occur when the number of defects is so large that half of the chain is in a ``defect state'' with respect to the other half. The ``active'' phase involves one atom excited in $\ket{2}$ (or $\ket{5}$), and $N/2-2$ in $\ket{0}$ ($\ket{1}$), which participate in two-level subradiance: the atoms exchange photons with $q=1(-1)$ and get excited and decay via closed transitions. The ``inactive'' or defect phase consists of $N/2+1$ atoms in $\ket{1}$ ($\ket{0}$). It should be noted that the eigenstate is a spatial superposition, where the active part can equally occupy the left and right side of the chain, with a domain wall in between the phases, as shown in Fig.~\ref{Fig5}(a). 

To observe signatures of phase separation, we calculate connected correlation functions between different Zeeman sublevels $m,n$. These are defined as
\begin{equation}
\mathcal{C}_{mn}(i,j)=\braket{\hat{\sigma}^i_{mm}\hat{\sigma}^j_{nn}}-\braket{\hat{\sigma}^i_{mm}}\braket{\hat{\sigma}^j_{nn}}
\end{equation}
where the expectation value is calculated over the most subradiant eigenstate, and $i,j$ are atomic indices. Figure~\ref{Fig5}(b) shows the correlations $\mathcal{C}_{00}(i,j)$ between $\ket{0}$-states in different locations for the most subradiant state of a chain of $N=14$ atoms. We observe $\ket{0}$-states clustering around each other, and ``repelling'' atoms in $\ket{1}$-states, displaying anti-correlations. The orange line in Fig.~\ref{Fig5}(c) shows the scaling of the decay rate with respect to atom number. Due to the exponentially large size of the Hilbert space, we can only exactly diagonalize up to $N=14$ atoms. However, based on our intuition of phase separation, we expect the decay rate to closely coincide with that obtained for the most subradiant eigenstate of a chain of $N/2-1$ atoms in the $|F^\text{max}_z|$ subspace. This calculation is plotted in Fig.~\ref{Fig5}(c) (red curve) and seems to agree with the scaling of the phase-separated states.

As in our previous analysis of other manifolds, thus far these results do not point to some fundamentally new mechanism for waveguiding: these states can be understood as the chain separating into two domains where ``two-level'' subradiance occurs independently from the other. 

\subsection{Symmetric states} 
We also numerically find evidence of another type of subradiant state, with significant population in states $\ket{3}$ and $\ket{4}$. These states also show no particular length scales or features in pairwise correlations, suggesting a qualitatively different mechanism for subradiance. To better grasp the underlying physics, we first introduce a ``toy model'' Hamiltonian, which is inspired by Eq.~\eqref{hq}. In particular, while the Hamiltonian can exactly be written in terms of pure spin wave operators for an infinite system, we consider a hypothetical \textit{finite} system whose Hamiltonian also takes the same form,
\begin{equation}\label{hq1}
H=\hbar\sum_{q=-1}^1\sum_{k}\left(J_{k,q}-\ii\frac{\Gamma_{k,q}}{2}\right)\hat{S}^{\dagger}_{k,q}\hat{S}_{k,q},
\end{equation}
where $\hat{S}^\dagger_{k,q}= N^{-1/2} \sum_{j}  e^{\ii k d j} \hat{\Sigma}^\dagger_{jq}$ and $J_{k,q}$ and $\Gamma_{k,q}$ are given in Eq.~\eqref{jjgg}. In particular, as these quantities correspond to the results for an infinite system, we have that $\Gamma_{k,q}=0$ when $|k|>k_0$. The wavevector $k$ is now taken to be a discrete index, with $k=2\pi n/N$ (with $n\in [0,N-1]$) to ensure periodic boundary conditions.

A perfectly subradiant eigenstate $\ket{\psi_{\tilde{k}}}$, of well-defined quasi-momentum $\tilde{k}$, with zero decay rate fulfills $\text{Im}\{H\ket{\psi_{\tilde{k}}}\}=0$. It should be noted that despite the apparent simplicity of Eq.~\eqref{hq1}, it is in general challenging to diagonalize, as the involved operators have complicated commutation relations. To proceed, we first find a state that fulfills the less demanding condition of being a zero-eigenvalue eigenstate of all spin operators $\hat{S}_{k,q}$ if $k\neq\tilde{k}$. Such a state is found to be
\begin{align}\nonumber
\ket{\psi_{\tilde{k}}}=\mathcal{N}\sum_{j=1}^N e^{\ii \tilde{k} z_j}\left(\ket{2_j}\ket{\mathcal{D}^j_{3/2}}+\beta\ket{3_j}\ket{\mathcal{D}^j_{1/2}}\right.\\\label{eig}
\left.\qquad{}+\beta\ket{4_j}\ket{\mathcal{D}^j_{-1/2}}+\ket{5_j}\ket{\mathcal{D}^j_{-3/2}}\right),
\end{align}
where $\beta=(C_{-1/2,1}/C_{1/2,1})$, $\mathcal{N}$ is a normalization constant, and 
\begin{equation}
\ket{\mathcal{D}^j_{\alpha}}=\sum_{m\neq j}\mathcal{P}_m\left\{\ket{0}^{\otimes n_0}\otimes\ket{1}^{\otimes n_1}\right\}
\end{equation}
is a Dicke state where all the atoms except $j$ are in ground states $\ket{0}$ or $\ket{1}$ such that the total angular momentum projection of the Dicke state is $\alpha=\{\pm1/2,\pm3/2\}$. In the above expression, $n_{0(1)}=(N-1\pM2\alpha)/2$ is the number of atoms in state $\ket{0}(\ket{1})$, and $\sum_m\mathcal{P}_m$ denotes the sum over all distinct permutations of the ground states. It should be noted that Dicke states are known to exhibit significant multipartite entanglement~\cite{T07,D11}. More importantly, the entanglement of this state cannot be obtained simply by applying a projection operator on a coherent state, as was the case for two-level subradiance. Remarkably, the limit where $H$ is composed of only one $k$-vector corresponds to Ref.~\cite{HKO17}, where they show the existence of states that are zeros of a single jump operator. Our states, however, must satisfy many relationships involving all spin wave operators $\hat{S}_{k\neq\tilde{k},q}$.

\begin{figure*}
\centerline{\includegraphics[width=\linewidth]{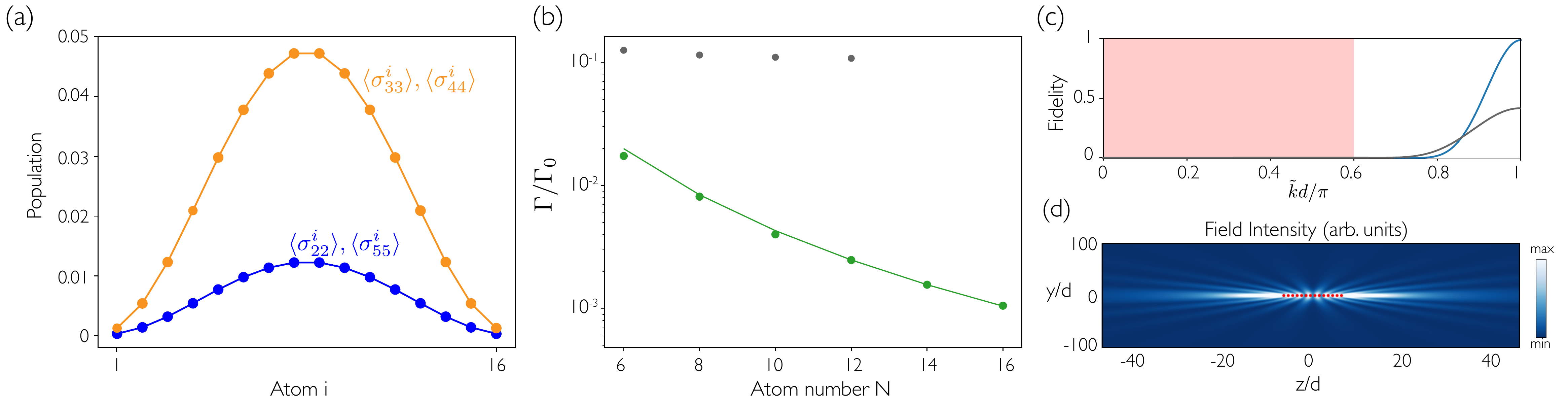}}
\caption{Symmetric states in the manifold of zero angular momentum projection ($F_z=0$). {\bf (a)} Spatial profile of the most subradiant eigenstate of a chain of $N=16$ atoms, where the $|q|=1$ component of the Green's function has been altered according to Eq.~\eqref{jshift}. {\bf (b)} Scaling of the decay rate as a function of atom number. The green dots correspond to the case where the $|q|=1$ component of the Green's function has been altered according to Eq.~\eqref{jshift}, and the gray dots are obtained for the unaltered case. The continuous line is a guide to the eye and scales as $\Gamma/\Gamma_0\sim1/N^3$. {\bf (c)} Overlap between the state $\ket{\psi_{\tilde{k}}}$ given by Eq.~\eqref{eig} and the most subradiant state of a finite chain obtained by numerical diagonalization, as a function of the wavevector $\tilde{k}$. The gray line is obtained for a chain of $N=12$ atoms in free space, while the blue line is obtained for $N=16$ atoms in a medium with dispersion engineering given by Eq.~\eqref{jshift}. The red shaded area represents the radiative wavevectors. {\bf (d)} Field intensity (arbitrary units) created by the most subradiant mode in a chain of $N=16$ atoms, obtained by including dispersion engineering. Red circles denote  atomic positions. For all plots, $d=0.3\lambda_0$. } \label{Fig6}
\end{figure*}

For the state in Eq.~\eqref{eig} to be a lossless eigenstate of $H$ it needs to fulfill 
\begin{equation}\label{eq16}
H\ket{\psi_{\tilde{k}}}=\hbar\sum_{q=-1}^1\left(J_{\tilde{k},q}-\ii\frac{\Gamma_{\tilde{k},q}}{2}\right)\hat{S}^{\dagger}_{\tilde{k},q}\hat{S}_{\tilde{k},q}\ket{\psi_{\tilde{k}}}=\lambda_{\tilde{k}}\ket{\psi_{\tilde{k}}},
\end{equation}
with $\lambda_{\tilde{k}} \in\mathbb{R}$. In the equation above, we have utilized the property that $S_{k\neq \tilde{k},q}\ket{\psi_{\tilde{k}}} = 0$, to eliminate all but a single spin-wave operator from $H\ket{\psi_{\tilde{k}}}$. For $|\tilde{k}|>k_0$ (such that $\Gamma_{\tilde{k},q}=0$), we find that the eigenstate equation~(\ref{eq16}) is satisfied provided that the dispersion relations for both polarizations $|q|={0,1}$ coincide at $\tilde{k}$, i.e. $J_{\tilde{k},q}\equiv J_{\tilde{k}}$, with a corresponding eigenvalue $\lambda_{\tilde{k}}=C_{-1/2,1}^2 \hbar J_{\tilde{k}}$. We subsequently show that an intersection of the dispersion relations at $|\tilde{k}|>k_0$ gives rise to waveguiding not only in our toy model, but in the original physical system.

In free space the dispersion relations  $J_{k,\pm1}$ and $J_{k,0}$ are in general different~\cite{AMA17}, and only intersect for a given $\tilde{k}>k_0$ for distances $d\lesssim0.17\lambda_0$. Numerically, however, it is difficult to confirm directly that an intersection in such a case leads to a decay rate approaching zero as $N\rightarrow\infty$. We attribute this to the fact that for the limited $N$ that we can simulate, the small lattice constant $d\ll\lambda_0$ still results in significant finite-size effects due to the short length of the chain, and because the intersection of the dispersion relations occurs close to radiative wavevectors $|\tilde{k}|\sim k_0$ (this is, for a short system, significant components of the many-body wavefunction still have wavevectors that couple to radiation). Thus, to better confirm our hypotheses, we add an additional short-range term to the $q=\pm 1$ interaction rates, $J_{ij,q=\pm1}\rightarrow J_{ijq=\pm1}+J'_{ijq=\pm1}$, given by 
\begin{align}\label{jshift}
J'_{ij,q=\pm 1}(\textbf{r})=-\frac{360 \Delta}{7} \left(\frac{d}{\pi r}\right)^4.
\end{align}
Here, $\Delta=J_{k=\pi/d,q=0}-J_{k=\pi/d,q=1}$ is the difference between the original free-space dispersion relations of Eq.~\eqref{jjgg}, evaluated at the edge of the Brillouin zone $k=\pi/d$.  It can readily be shown that this extra term guarantees that both the dispersion relations and the slopes of the $q=0, \pm 1$ polarizations coincide at the Brillouin zone edges, $|k|=\pi/d$. In the following, we will not focus on the details of how such dispersion engineering is implemented, although we note that it can potentially be realized by introducing some dielectric structure to change the Green's function itself, or by dressing of atomic levels.

Figure~\ref{Fig6}(a) shows the spatial profile of the most subradiant eigenstate of a finite chain of $N=16$ atoms obtained with dispersion engineering. The population in levels $\ket{2}$ and $\ket{5}$ is identical, and so are the populations in $\ket{3}$ and $\ket{4}$. In Fig.~\ref{Fig6}(b), we plot the decay rate of the most subradiant state vs. atom number. In particular, we observe a scaling of the decay rate with atom number of $\Gamma/\Gamma_0\sim 1/N^3$, identical to that observed for ``classical'' waveguiding. Figure~\ref{Fig6}(c) shows the fidelity between the most subradiant eigenstate of the finite chain and the toy-model state $\ket{\psi_{\tilde{k}}}$ given by Eq.~\eqref{eig}. There is extremely good agreement at  $\tilde{k}=\pi/d$. In particular the infidelity at $\tilde{k}=\pi/d$ scales as $\epsilon=1-|\braket{\psi|\psi_{\tilde{k}}}|^2\sim1/N^2$, which indicates that our toy model accurately captures the physics of the actual system. Finally, Fig.~\ref{Fig6}(d) shows the field intensity emitted by the most subradiant eigenstate of the chain with dispersion engineering. The intensity pattern is consistent with waveguiding along the chain, where most of the field is radiated through the edges. We note that these states are robust against classical fluctuations in the atomic positions, as shown in the SI.

We emphasize the importance of realizing an intersection of the dispersion relations $J_{k,\pm 1}$ and $J_{k,0}$, in order to yield this new class of highly entangled, waveguiding states. In particular, in the gray points and curves of Figs.~\ref{Fig6}(b) and (c), respectively, we plot the corresponding results where only the free-space Green's function is used, without the additional term of Eq.~\eqref{jshift}. It can be seen that although a moderate reduction of decay rate $\Gamma/\Gamma_0\sim0.1$ can be achieved, this rate apparently does not decrease with increasing $N$. Furthermore, the overlap fidelity between the numerically obtained eigenstate and the ansatz state does not approach 1, as the differing coherent interactions associated with $q=0,\pm 1$ mix in additional contributions to the eigenstate. We note that, in contrast with the dispersion-engineered situation, these are not the most subradiant eigenstates in the $F_z=0$ manifold (as they are given by the phase-separated states). Thus, we have found these states by exactly diagonalizing the Hamiltonian and filtering out states such that $\sum_j\braket{\sigma^j_{33}}>\sum_j\braket{\sigma^j_{22}}$, and then selecting the most subradiant states that satisfy that condition. In this situation, we need to diagonalize the full Hamiltonian matrix, instead of simply finding the eigenvalue with the smallest decay rate by using iterative sparse-matrix diagonalization algorithms, which limits the size of the chain that we can successfully diagonalize to $N=12$.

\section{Discussion}
To summarize, we have shown that arrays of atoms with hyperfine structure can support highly subradiant, waveguiding states, and we have elucidated the conditions for their existence. In contrast to the ``classical'' effect that occurs in atoms with a single ground state, here, the waveguiding is fundamentally enabled by rich, many-body correlations within the ground-state manifold.

Having shown the existence of true waveguiding states, one interesting question going forward is how novel phenomena or applications, previously identified for simple atoms, can be encoded into these highly entangled states. For example, it would be interesting to investigate if such states support more powerful quantum memory protocols, or if generalizations of such states exist in higher dimensions, such as to support topological edge states. In the case of higher dimensions, while numerics will likely be highly challenging, a promising approach could be the generalization of toy models such as~Eq.~\eqref{hq1}. It would also be interesting to investigate multi-excitation subradiant states, e.g., to see if they exhibit the same ``fermionic'' correlations as multi-excitation states in atoms with simple (two-level) structure~\cite{AMA17}.

An important associated question is how the necessary many-body entanglement in the ground-state manifold can be generated in the first place. We speculate that the same correlated dissipation processes encoded in the dipole-dipole interactions of Eq.~\eqref{fullham} might be used to generate the necessary entanglement, such as through the steady state obtained under constant driving (i.e., correlated optical pumping).

More generally, the insight developed in this work could give rise to broader opportunities. For example, under constant driving, one could investigate whether correlated dissipation can give rise to useful many-body correlations within the ground-state manifold, such as for quantum-enhanced metrology~\cite{LSV10,WJK10,BCN14,HEK16,MP19}. It would also be interesting to more systematically understand the forms and range of entanglement that can arise, as the Green's function is varied (e.g. through a dielectric structure) and/or the atomic level structure is altered. Finally, as an optical phenomenon, it is intriguing that we have identified a mechanism for waveguiding that explicitly relies on entanglement. It would be interesting to more broadly search for novel mechanisms of waveguiding and other optical effects~\cite{BPP19}, which can only emerge through quantum correlations.

\vspace{10pt}
\textbf{Acknowledgments.}-- A. A.-G. work was supported as part of Programmable Quantum Materials, an Energy Frontier Research Center funded by the U.S. Department of Energy (DOE), Office of Science, Basic Energy Sciences (BES), under award DE-SC0019443. D.E.C. acknowledges support from Fundaci\'on Ramon Areces, Fundacio Privada Cellex, Spanish MINECO Severo Ochoa Program SEV-2015-0522, Spanish Plan Nacional Grant ALIQS (funded by the MCIU, AEI, and FEDER), CERCA Programme/Generalitat de Catalunya, ERC Starting Grant FOQAL, and AGAUR Grant 2017 SGR 1334. H.J.K. acknowledges funding from the Office of Naval Research (ONR) Grant N00014-16-1-2399, the ONR Multidisciplinary University Research Initiative (MURI) Quantum Opto-Mechanics with Atoms and Nanostructured Diamond Grant N00014-15-1-2761, and the Air Force Office of Scientific Research MURI Photonic Quantum Matter Grant FA9550-16-1-0323.

\bibliography{refs_jan_2019.bib}

\clearpage
\onecolumngrid

\appendix

\section*{Supplementary Information}
Here we present further details on the physics of defect states and we show that subradiant states are robust against (classical) fluctuations in atomic positions. 

\section{Defect states}
\begin{figure*}[h]
\centerline{\includegraphics[width=\linewidth]{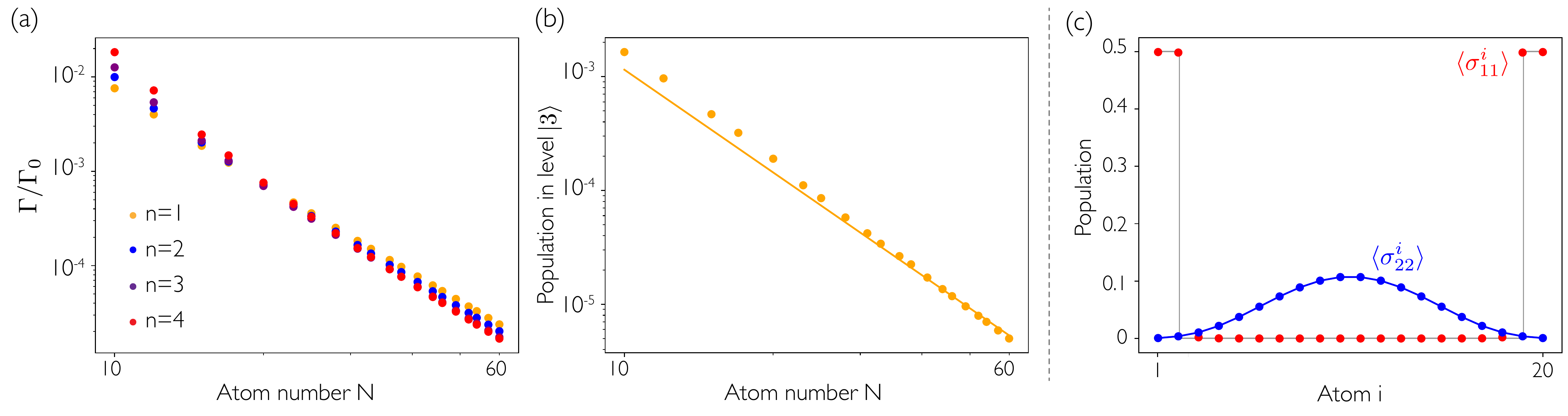}}
\caption{Details of defect states. \textbf{(a)} Scaling of the most subradiant eigenstate with a defect in atom in position $n$ vs. atom number, for $|F_z|=|F_z^\text{max}|-1$. \textbf{(b)} Total population in level $\ket{3}$ vs. atom number $N$, for the most subradiant state in $|F_z|=|F_z^\text{max}|-1$. The continuous line is a guide to the eye showing a scaling of $\sum_j\braket{\sigma_{33}^j}\sim1/N^3$. \textbf{(c)} Spatial profile of the populations of levels $\ket{1}$ (red) and $\ket{2}$ (blue) in the most subradiant eigenstate of a chain of $N=20$ atoms, for $|F_z|=|F_z^\text{max}|-2$. The two defects accumulate at the edge of the chain. The population of $\ket{3}$ is negligible in all atoms. The continuous lines are guides to the eye. For all plots, $d=0.3\lambda_0$.}\label{FigSI}
\end{figure*}
In Fig.~\ref{FigSI}(a) we plot the decay rate of the most subradiant state in the manifold $|F_z|=|F_z^\text{max}|-1$ as a function of atom number $N$, and classified by the position of the defect atom ($n=1,2,3,4$). It can be seen that although the position of the defect atom $n$ changes for the most subradiant state overall (minimized over all possible positions $n$), all of these defect states exhibit qualitatively similar behavior. Considering the most subradiant defect state, we plot in Fig.~\ref{FigSI}(b) the total population of state $\ket{3}$, which is seen to decrease with atom number as $\sum_j\braket{\hat{\sigma}_{33}^j}\sim 1/N^3$. In manifolds with lower angular momentum, defect atoms pile up at the edges of the chain. This is illustrated in Fig.~\ref{FigSI}(c), where we plot the individual atomic state populations for the most subradiant state of $N=20$ atoms, in the $|F_z|=|F_z^\text{max}|-2$ manifold. Here, the two defect atoms, characterized by a large population in state $\ket{1}$, are positioned as the two outermost atoms at either of the chain ends. 

\section{Effect of disorder}

\begin{figure*}
\centerline{\includegraphics[width=\linewidth]{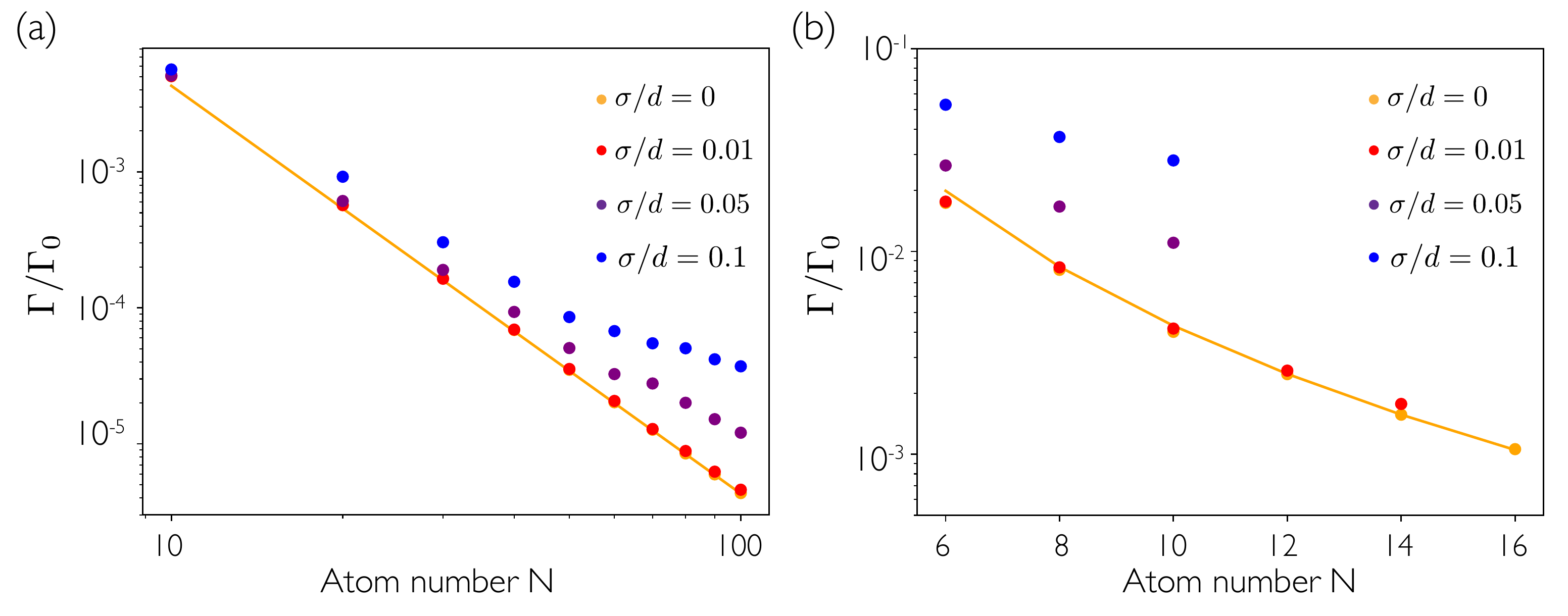}}
\caption{Influence of classical position disorder $\sigma$ in the decay rate of the most subradiant state. Averaged scaling of the most subradiant eigenstate vs. atom number for: \textbf{(a)}  $|F_z|=|F_z^\text{max}|$, and \textbf{(b)} $|F_z|=0$ (symmetric states). The continuous lines are guides to the eye showing a scaling of $\Gamma/\Gamma_0\sim 1/N^3$. The disorder is assumed to be following a normal distribution of standard deviation $\sigma$. We consider 200 random configurations for each value of $\sigma$. For all plots, $d=0.3\lambda_0$. }\label{FigSId}
\end{figure*}

In Fig.~\ref{FigSId} we plot the decay rate of the most subradiant state in the (a) $|F_z|=|F_z^\text{max}|$ and (b) $|F_z|=0$ manifolds as a function of atom number $N$ for different degrees of spatial disorder. Specifically, here we consider classical disorder, where each atom is randomly displaced around its equilibrium position by a normal distribution of standard deviation $\sigma$. We plot the decay rate of the most subradiant eigenstate after performing an average over disorder realizations. The chain lies along $z$ and the disorder is introduced only along this dimension (i.e. there is no disorder in the positions along $x,y$), as disorder in the perpendicular directions would break the conservation of angular momentum projection, which would make the calculations unmanageable. While increasing disorder leads to an average increase in the decay rate, it is clear from the figure that there are still deeply subradiant states even in the presence of significant fluctuations ($\sigma/d=0.1$). The number of disorder realizations is 200. For (b), the numerical difficulty of the problem prevents us from going beyond $N=10-12$ for significant disorder. In particular, sparse diagonalization methods seem to fail, which requires one to resort to exact diagonalization instead. Nevertheless, there does not seem to be any drastic difference in how it is affected by disorder, as compared to classical subradiance.

We now briefly discuss the case of quantum disorder. Specifically, we consider the that each atom is tightly trapped in the Lamb-Dicke regime, and in the motional ground state of its trap. When an atom is excited, there is the possibility of an inelastic process, where a photon is emitted and a phonon is created at the same time. As shown in Ref.~\cite{GGV19}, this leads to an additional decay rate of each atom of $\Gamma'\sim \Gamma_0 \eta^2$, where $\eta$ is the Lamb-Dicke parameter. This decay process is independent (does not depend on correlations), as the final atomic motional state (with one phonon) is distinguishable.


\end{document}